\documentclass[times]{elsarticle}
\usepackage{graphicx}
\makeatletter
\def\ps@pprintTitle{%
 \let\@oddhead\@empty
 \let\@evenhead\@empty
 \def\@oddfoot{}%
 \let\@evenfoot\@oddfoot}

\usepackage{etoolbox}

\usepackage{amssymb,amsfonts,amsmath}
\usepackage{lineno}

\begin{document}

\begin{frontmatter}

\title{{\bf Efficient conversion of chemical energy into mechanical work by Hsp70 chaperones}}

\author[add1,add2,*]{S. Assenza}
\author[add3,add4,*]{A. S. Sassi}
\author[add5]{R. Kellner}
\author[add5,add6]{B. Schuler}
\author[add3,add7]{P. De Los Rios}
\author[add8]{A. Barducci}
\ead{alessandro.barducci@cbs.cnrs.fr}

\address[add1]{Laboratory of Food and Soft Materials, ETH Z\"urich, CH-8092 Z\"urich, Switzerland}
\address[add2]{Departamento de F\'isica Te\'orica de la Materia Condensada, Universidad Aut\'onoma de Madrid, E-28049 Madrid, Spain}
\address[add3]{Institute of Physics, School of Basic Sciences, Ecole Polytechnique F\' ed\' erale de Lausanne (EPFL), CH-1015 Lausanne, Switzerland}
\address[add4]{IBM T. J. Watson Research Center, Yorktown Heights, New York, United States of America}
\address[add5]{Department of Biochemistry, University of Zurich, CH-8057 Zurich, Switzerland}
\address[add6]{Department of Physics, University of Zurich, CH-8057 Zurich, Switzerland}
\address[add7]{Institute of Bioengineering, School of Life Sciences, Ecole Polytechnique F\'ed\'erale de Lausanne (EPFL), CH-1015 Lausanne, Switzerland}
\address[add8]{Centre de Biochimie Structurale (CBS), INSERM, CNRS, Universit\' e de Montpellier, Montpellier, France}
\address[*]{These two authors contributed equally}

\begin{abstract}
Hsp70 molecular chaperones are abundant ATP-dependent nanomachines that actively reshape non-native, misfolded proteins and assist a wide variety of essential cellular processes. Here we combine complementary computational/theoretical approaches to elucidate the structural and thermodynamic details of the chaperone-induced expansion of a substrate protein, with a particular emphasis on the critical role played by ATP hydrolysis. We first determine the conformational free-energy cost of the substrate expansion due to the binding of multiple chaperones using coarse-grained molecular simulations. We then exploit this result to implement a non-equilibrium rate model which estimates the degree of  expansion as a function of the free energy provided by ATP hydrolysis. Our results are in quantitative agreement with recent single-molecule FRET experiments and highlight the stark non-equilibrium nature of the process, showing that Hsp70s are optimized to convert effectively chemical energy into mechanical work close to physiological conditions.
\end{abstract}


\end{frontmatter}

\section{Introduction}
Even though \textit{in vitro} most proteins can reach their native structure spontaneously\cite{Anfinsen}, this is not always the case in cellular conditions and proteins can populate misfolded states which can form cytotoxic aggregates\cite{Dobson}. In order to counteract misfolding and aggregation, cells employ specialized proteins, called {\em molecular chaperones}, which act on non-native protein substrates by processes that stringently depend on ATP hydrolysis for most chaperone families\cite{mayer,hartl}. Among them, the ubiquitous 70 kDa heat shock proteins (Hsp70s) play a special role because they assist a plethora of fundamental cellular processes beyond prevention of aggregation.

Hsp70s consist of  two domains \cite{mayer,gierasch}. The substrate binding domain (SBD) interacts with disparate substrate proteins, whereas the nucleotide binding domain (NBD) is responsible for the binding and hydrolysis of ATP. The two domains are allosterically coupled, and the nature of the nucleotide bound to the NBD affects the structure of the SBD and as a consequence the affinity for the substrate and its association/dissociation rates. More precisely, the chaperone in the ATP-bound state is characterized by binding and unbinding rates that are orders of magnitude larger than those measured when ADP is bound \cite{Bukau2000}. Furthermore the coupling is bidirectional: the substrate, together with a co-localized J-domain protein (JDP) that serves as cochaperone, greatly accelerates the hydrolysis of ATP. Substrate binding thus benefits from the fast association rate of the ATP-bound state and the slow dissociation rate of the ADP-bound state, resulting in a non-equilibrium affinity (\textit{ultra-affinity}) that can be enhanced beyond the limits imposed by thermodynamic equilibrium \cite{Bard_ultra}.

Several lines of evidence suggest that the binding of Hsp70s to a polypeptide induces its expansion. Nuclear Magnetic Resonance (NMR) measurements have shown that Hsp70s destabilize the tertiary structure of several different substrates \cite{Cavagnero,Kay}. Biochemical assays revealed that binding of Hsp70 increases the sensitivity of misfolded Luciferase to proteolysis and decreases its propensity to bind Thioflavin-T, strongly suggesting a loss of compactness \cite{DeLos_2010}. Moreover, a single-molecule study based on F\"orster resonance energy transfer (FRET) spectroscopy 
quantified the considerable expansion of unfolded rhodanese in native conditions upon binding of multiple Hsp70 chaperones \cite{Schuler2014}. In particular, this study revealed that the expansion is stringently ATP-dependent, because upon ATP exhaustion the system relaxes to the expansion values observed in the absence of chaperones \cite{Schuler2014}. Building on this result, we elucidate here the coupling between the expansion of the substrate and the external energy source provided by ATP hydrolysis. To this aim, we first explore the structural and energetic features of Hsp70-bound rhodanese using Molecular Dynamics (MD) simulations. We next integrate this molecular information into a rate model that explicitly includes the Hsp70-rhodanese interactions and the chaperone ATPase cycle, thus clarifying the role played by energy consumption in the expansion of the substrate.

\section{Results}
\subsection{Structural and thermodynamic characterisation of chaperone-substrate complexes}  
\begin{figure*}
\begin{center}
\includegraphics[width=\linewidth]{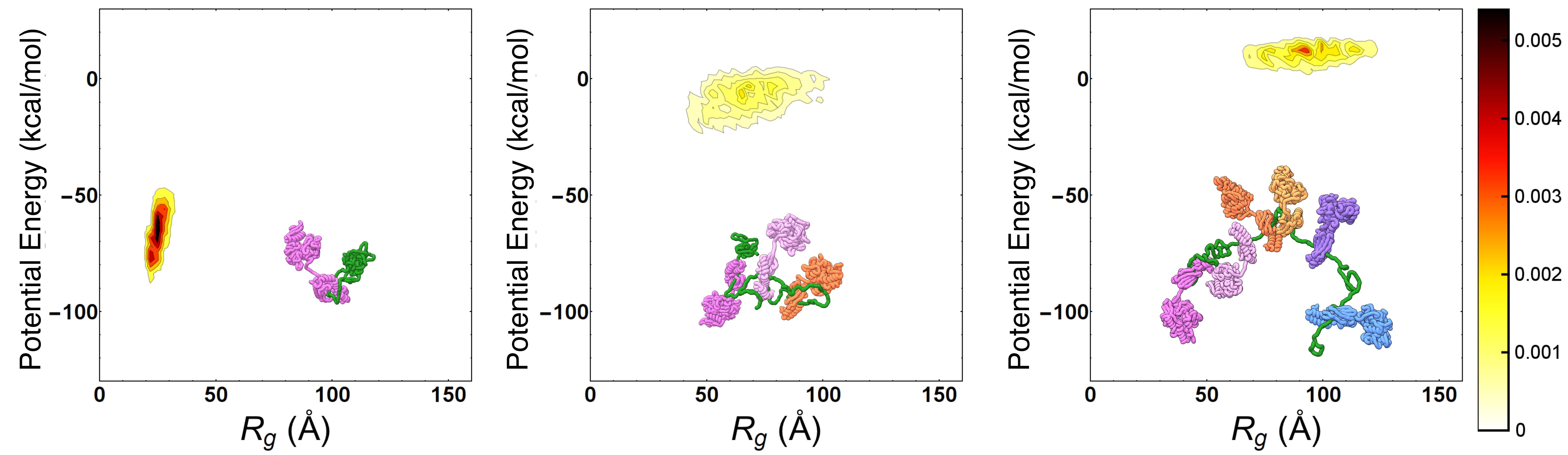}
\caption{\label{fig:histoRg_DG} 
Probability density maps of substrate potential energy and radius of gyration for representative Hsp70/rhodanese complexes with one (left), three (center) and six (right) bound chaperones. The different Hsp70 chaperones have been represented with different colors to ease their discernibility.
}
\end{center}
\end{figure*}

To characterize the main features of chaperone-induced expansion, we performed MD simulations of the Hsp70/rhodanese complexes. We relied on a one-bead-per-residue Coarse Grained (CG) force field \cite{smith2014calibrated}, which  has been tailored to match experimental FRET data of intrinsically-disordered proteins and satisfactorily reproduce the compactness of unfolded rhodanese in native conditions without any further tuning (see SI). In particular, we focused on hydrophobic and excluded volume interactions while neglecting the electrostatic contribution which is negligible in rhodanese and plays only a minor role in Hsp70/rhodanese complexes according to FRET experiments (see SI).
Hsp70 chaperones were modeled with a structure-based potential built upon the ADP-bound conformation, and restrained on binding sites on the substrate. 
 We identified six binding sites on the rhodanese sequence using two distinct bioinformatic algorithms\cite{vanDurme,Rudiger}. Considering that each binding site could be either free or bound to a Hsp70 protein, we thus took into account a total of $2^6=64$ distinct chaperone/substrate complexes, which were exhaustively simulated.
 In Fig.\ref{fig:histoRg_DG} we report the distributions of the substrate potential energy and of the radius of gyration ($R_g$) for three representative complexes with one (left), three (center) and six (right) bound chaperones. As previously noticed\cite{Schuler2014}, chaperone binding leads to larger radii of gyration and higher potential energies, implying that the excluded volume interactions due to the large Hsp70s progressively expand the complex and disrupt the attractive intra-chain interactions in rhodanese. 

We then calculated the conformational free energy of all the possible chaperone/rhodanese complexes to obtain a quantitative picture of the energy landscape governing the chaperone-induced expansion. To this aim, we performed extensive sets of non-equilibrium steering MD trajectories for each complex, and measured the work needed to steer it to a completely extended reference structure ($R_{gyr}>260\mbox{\AA}$), whose conformational free-energy is not affected by chaperone binding. Equilibrium free-energy differences with respect to this reference state were then estimated from non-equilibrium work distributions via the Jarzynski equality\cite{jarzynski1997nonequilibrium}, thus allowing the determination of the conformational free-energy $\Delta G$ of each distinct chaperone/substrate complex. 

\begin{figure}[h] 
 \begin{center}
 \includegraphics[width=\linewidth]{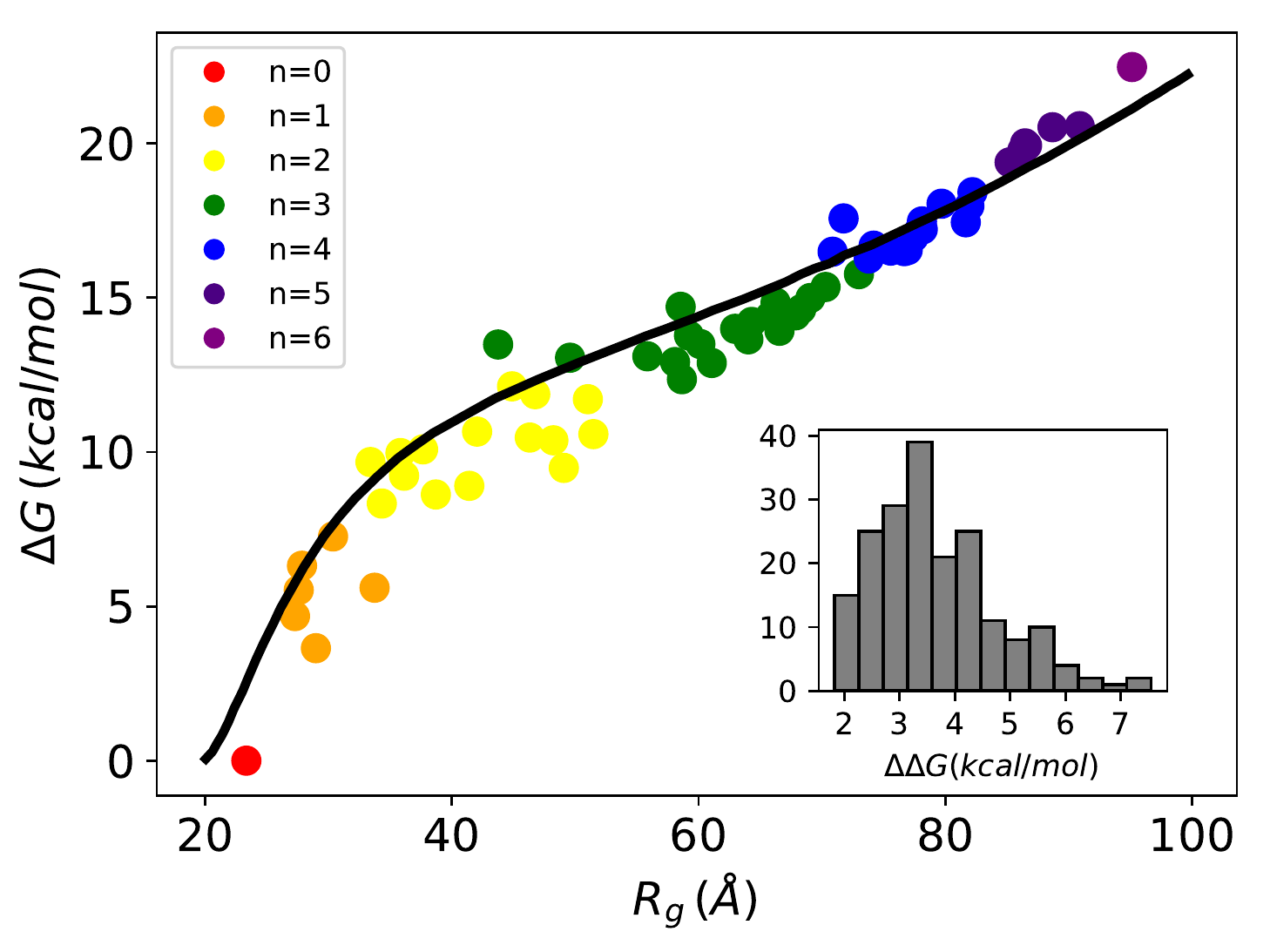}
 \caption{\label{fig:histo_Kd} 
Free energy $\Delta G$  as a function of the radius of gyration $R_g$ for all the 64 possible binding configurations. Different colors represent different numbers of chaperones bound. The black curve was obtained using the model in \cite{Sanchez} (see SI). (inset) Histogram of the free energy cost $\Delta\Delta G$ of the binding of an additional chaperone.}

 \end{center}
 \end{figure}

In Fig.\ref{fig:histo_Kd} (main) we report $\Delta G$ for each complex as a function of its mean radius of gyration using different colors for different stoichiometries. The conformational free energy increased with the swelling of the substrate due to the progressive binding of the chaperones. The increase in substrate potential energy due to the loss of intra-chain interactions upon Hsp70 binding is therefore only marginally compensated by the gain in conformational entropy. Notably, the conformational free-energy is not uniquely determined by the stoichiometry, and is significantly affected by the specific binding pattern. The conformational free-energy cost $\Delta\Delta G$ of adding a single chaperone (inset in Fig.\ref{fig:histo_Kd}) is positive for all complexes, but it varies from 2 kcal/mol up to 7 kcal/mol depending on the stoichiometry of the complex and on the particular choice of the binding sites.
The increase of $\Delta G$ as a function of $R_g$ is quantitatively captured by Sanchez theory (\cite{Sanchez} and SI) for the coil-to-globule collapse transition in polymers (see Fig.\ref{fig:histo_Kd}). Remarkably, the excellent agreement is not the outcome of a fitting procedure since all the parameters were extracted from experiments (see Methods). This result further reinforces the reliability of our simulations as well as the general applicability of the present setup beyond the particular system considered in this work.  

\subsection{ATP hydrolysis promotes multiple chaperone binding}
The structural and thermodynamic characterization obtained by molecular simulations can be profitably complemented by a kinetic model encompassing relevant biochemical processes
in order to determine the probability of each chaperone/substrate complex as a function of the chemical conditions. Notably, a model of the Hsp70 biochemical cycle based on experimental rates was previously used to illustrate how ATP-hydrolysis may result into non-equilibrium ultra-affinity for peptide substrates \cite{Bard_ultra}. Here we extend this result to the more complex case of Hsp70-induced expansion by taking into account multiple chaperone binding events and their consequences on the conformational free energy of the substrate. 

 \begin{figure}[h]
 \begin{center}
 \includegraphics[width=0.75\linewidth]{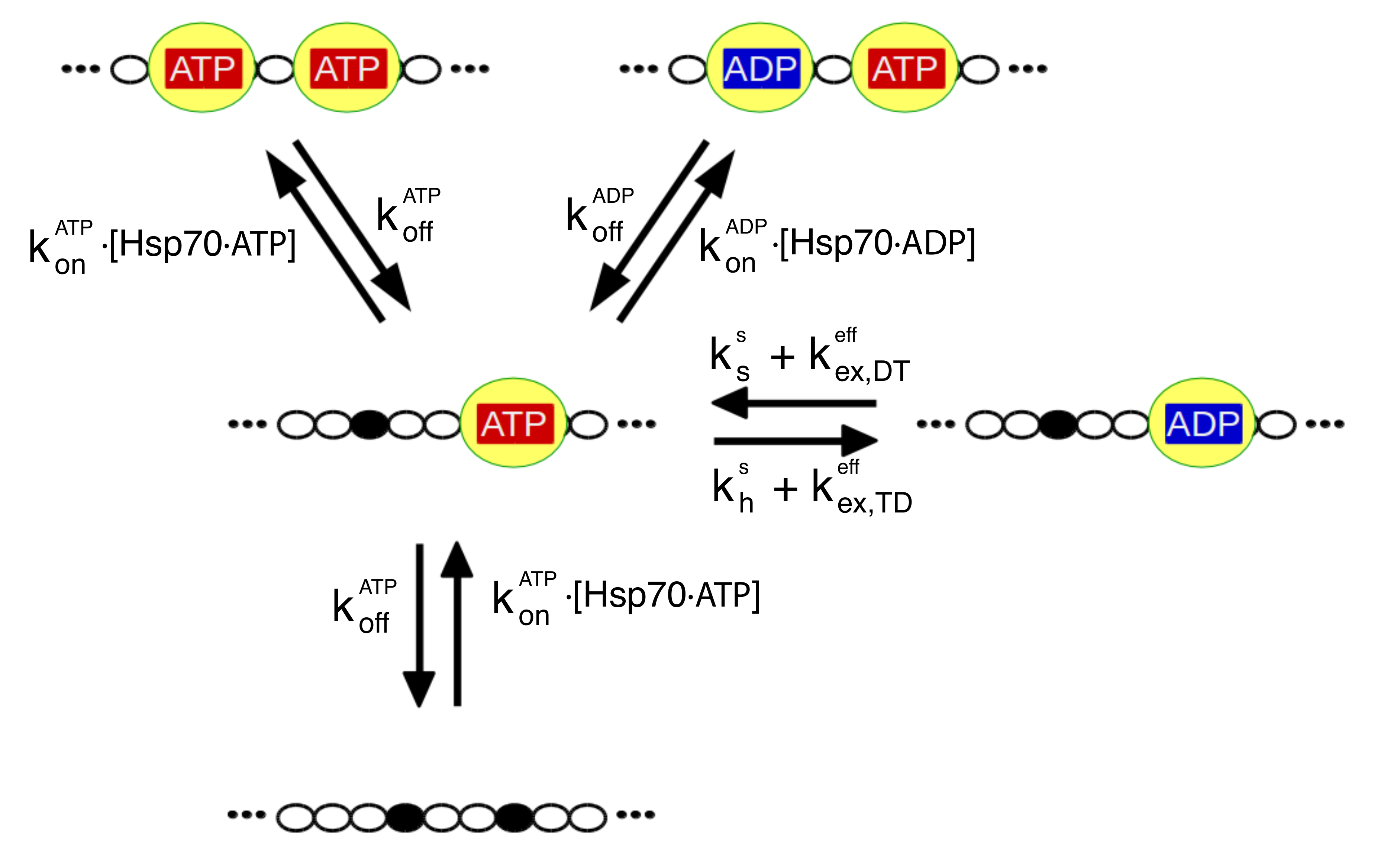}
 \caption{\label{fig:scheme} 
Portion of the biochemical cycle. Each binding site of rhodanese (filled in black) can be either free or occupied by an Hsp70 (yellow), which in turn can be either ADP- or ATP-bound.
The rate of each reaction is highlighted, as detailed in the SI. The binding constant, $k_{on}^{ATP}$ and $k_{on}^{ADP}$, are computed according to (\ref{binding constant}).
 }
 \end{center}
 \end{figure}
 
In our model each state corresponds to a single configuration of the chaperone/substrate complex, which is defined by the occupation state of the six Hsp70 binding sites on rhodanese. Each site can be either free or occupied by an ADP- or ATP-bound chaperone for a total of $3^6=729$ different states. All the relevant molecular processes corresponding to transitions between these states are explicitly modeled, including chaperone binding/unbinding, nucleotide exchange and ATP hydrolysis (see Fig.\ref{fig:scheme}).  We took advantage of available biochemical data for determining the rate constants associated to all the relevant reactions (see methods and SI). Importantly, kinetic rates for Hsp70 binding were modulated by the conformational free-energies determined by CG MD simulations. 
Indeed, the unbinding rates of Hsp70 from large-sized protein substrates were observed to be similar to the ones from small peptides, whereas the binding rates can be up to two orders of magnitude smaller\cite{Schuler2014,Bukau2000,Schmid}. This evidence was further corroborated by a recent NMR study\cite{Kay} suggesting a conformational selection scenario where the energetic cost due to substrate expansion mostly affects the Hsp70/rhodanese binding rate.
Following the experiments, we thus considered a substrate-independent unbinding rate constant $k_{off}$, while we expressed the binding rate constant as  
\begin{equation}
k_{on, ij}=k^0_{on}\exp[-\beta\Delta\Delta G_{ij}]\,,
\label{binding constant}
\end{equation} 
where $\beta=1/k_bT$, $k^0_{on}$ is the binding rate measured for a peptide substrate, and  $\Delta\Delta G_{ij}$ is the conformational free-energy cost of Hsp70 binding, which depends on the specific initial and final binding patterns $i$ and $j$ in the rhodanese/chaperone complex (see Fig.\ref{fig:histo_Kd}, inset). The interactions with JDP cochaperones were not explicitly modeled but the cochaperones were assumed to be colocalized with the substrate, so that their effect was implicitly taken into account in the choice of the rate constants for the ATP hydrolysis \cite{Tomita,Kampinga}. 

\begin{figure}[h]
 \begin{center}
 \includegraphics[width=\linewidth]{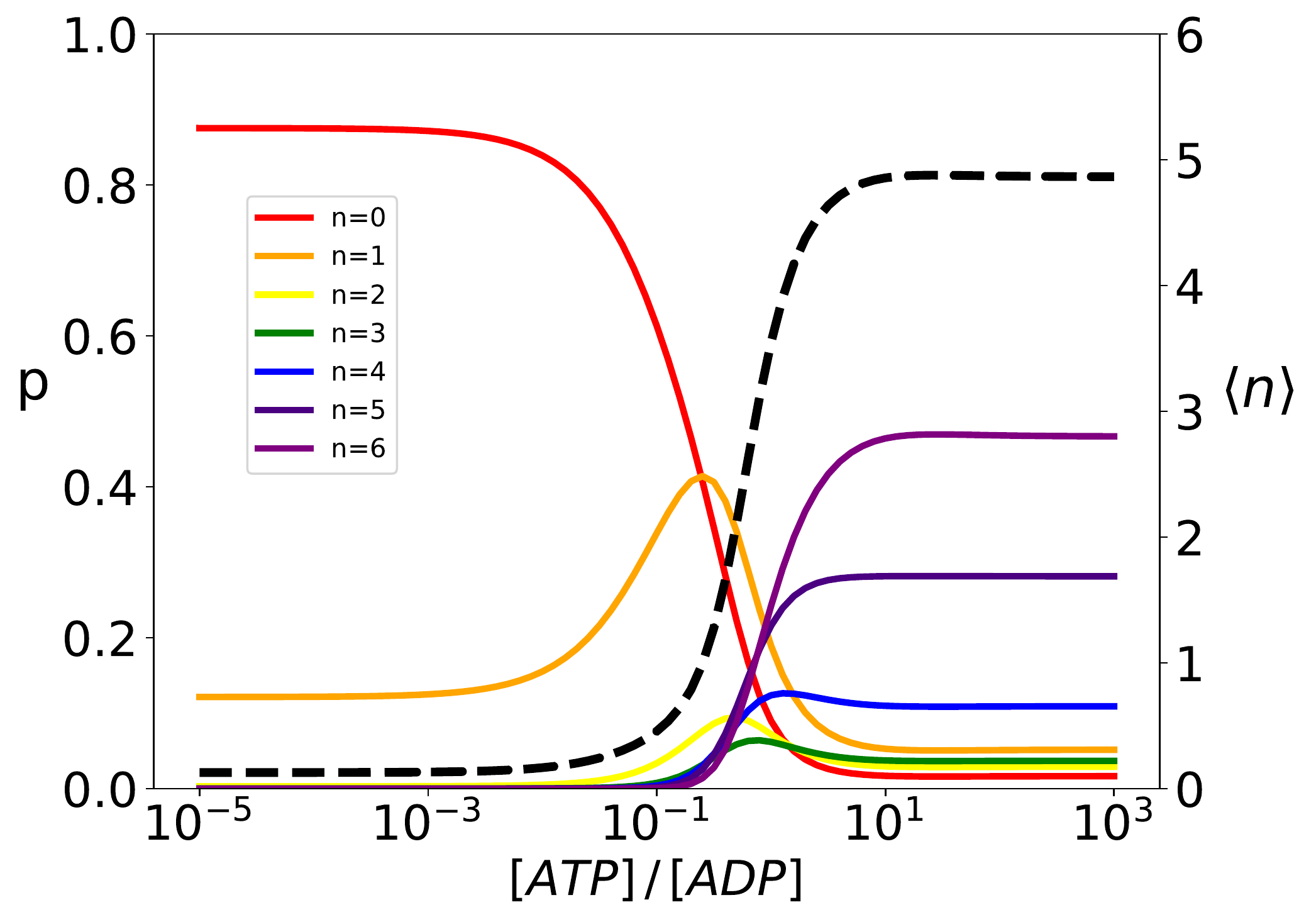}
 \caption{\label{fig:fig4} 
Probabilities of the state with $n$ chaperones bound as a function of [ATP]/[ADP], for different stoichiometries $n$. The dashed line indicates the mean value $\left<n\right>$.
 }
\end{center}
\end{figure}

The analytical solution of the model provides the steady-state probability of each binding configuration and allows the exploration of their dependence on the biochemical parameters. It is particularly instructive to investigate the system behavior as a function of the ratio between the concentration of ATP and ADP, which is intimately connected to the energy released by ATP hydrolysis. At thermodynamic equilibrium, the $[ATP]/[ADP]$ ratio is greatly tilted in favor of ADP  ($[ATP]_{eq}/[ADP]_{eq} \simeq 10^{-9}-10^{-8}$, \cite{Alberty}) whereas in the cell ATP is maintained in excess over ADP by energy-consuming chemostats ($[ATP]/[ADP] > 1$, \cite{Milo2016}). The $[ATP]/[ADP]$ ratio hence determines how far the system is from equilibrium, thus representing a natural control parameter for the non-equilibrium biochemical cycle. We thus report in Fig.\ref{fig:fig4} the compound probabilities for complexes with the same stoichiometry $n$ as a function of this nucleotide ratio. In conditions close to equilibrium (very low values of $[ATP]/[ADP]$), the vast majority of the substrate proteins are free and only about 10\% of them are bound to a single chaperone. The population of equimolar complexes increases for 
 $[ATP]/[ADP]$ between $10^{-2}$ and $10^{-1}$ and gives way to larger complexes with multiple chaperones for higher values of the nucleotide ratio.

For $[ATP]/[ADP]>1$, most substrates are bound to at least $4$ chaperones, with an average stoichiometry $\left<n\right>\sim 4.9$. Further increase of the nucleotide ratio does not significantly change this scenario indicating an almost constant behaviour in large excess of ATP $([ATP]/[ADP]>10)$. 

\subsection{Substrate expansion as a non-equilibrium process}
Combining the steady-state probabilities derived from the rate model with the results of the MD simulations, we can now exhaustively characterize the structural properties of the system. This provides the opportunity to directly compare our model with the results from FRET experiments both in equilibrium and non-equilibrium conditions.
To this aim, we first focused on the average radius of gyration of the system at thermodynamic equilibrium  ($[ATP]\ll[ADP]$)  or in non-equilibrium conditions with ATP in large excess over ADP ($[ATP]/[ADP]>10$). 

\begin{figure}[h]
\begin{center}
\includegraphics[width=0.97\linewidth]{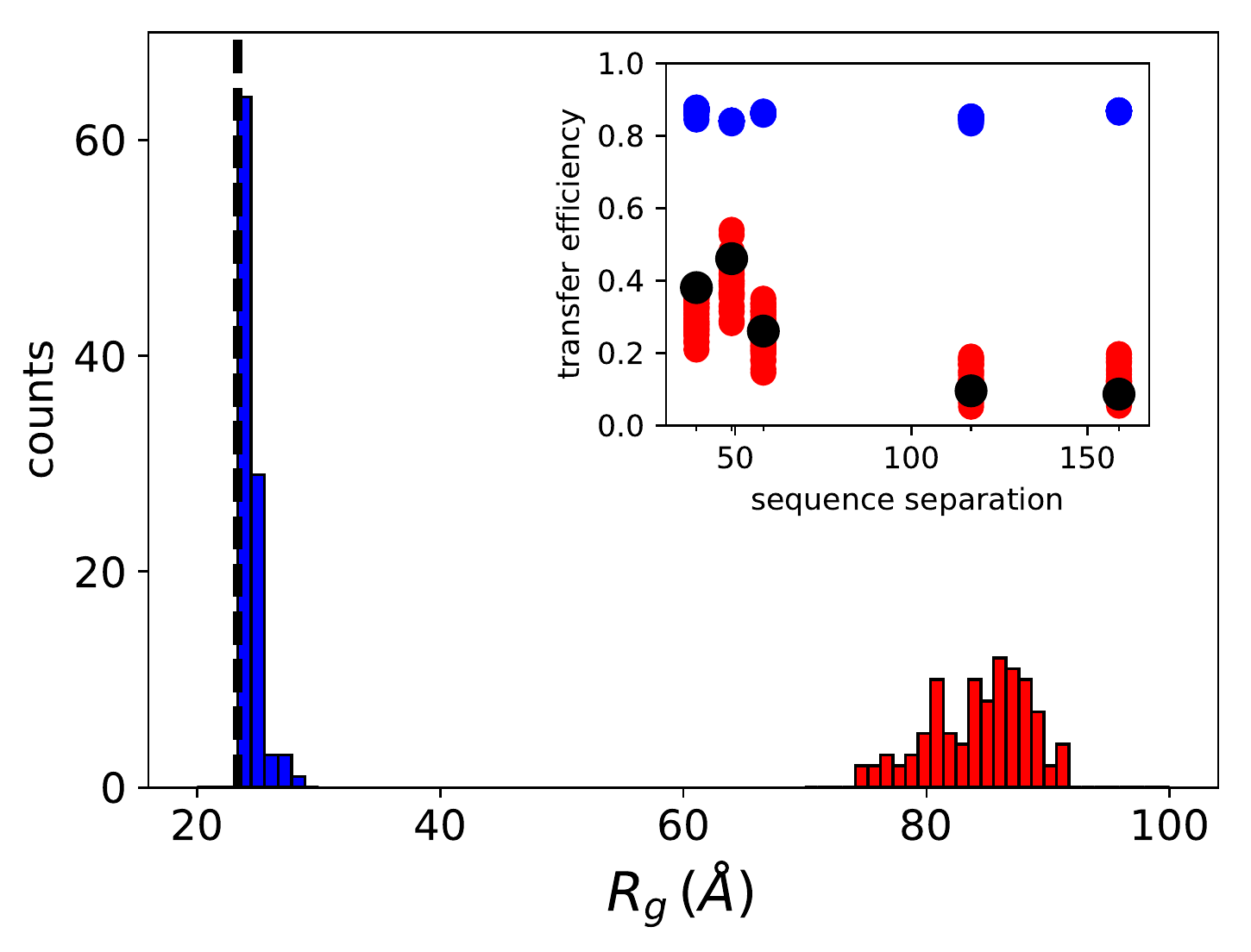}
\caption{\label{fig:fig5} 
Histogram of the radius of gyration for equilibrium (blue) and non equilibrium (red) values of [ATP]. The black dashed line indicates the average radius of unbound rhodaneses. (inset) FRET transfer efficiencies as a function of the sequence separation between the fluorescent dyes. The black circles correspond to the experimental values \cite{Schuler2014}. Calculated efficiencies taking into account uncertainties are reported as blue (equilibrium conditions) and red circles (ATP excess)}
 \end{center}
 \end{figure}

We also  probed the robustness of this result with respect to inaccuracies in the molecular model by taking into account normally-distributed errors on the conformational free energies $\Delta G_i$. The results are reported as histograms in Fig.\ref{fig:fig5} (top) and they suggest that at equilibrium the average radius of gyration is extremely close to what would be measured in the case of free substrate (dashed line). This is in agreement with the experimental observation that Hsp70 cannot significantly associate to rhodanese in these conditions\cite{Schuler2014}. Conversely, in large excess of ATP we observe a substantial swelling of the substrate ($75<R_g<95$\AA) due to the ultra-affine binding of Hsp70s. This finding is fully compatible with the size of DnaK/DnaJ/rhodanese complexes determined by sm-FRET experiments in excess of ATP \cite{Schuler2014}. In this regime, the limited effects of cochaperone binding on substrate conformations, which are not explicitly included in the model, play a minor role in determining the global expansion of the complex.

A more quantitative comparison between the model and the FRET results can be achieved by back-calculating the transfer efficiencies that were experimentally measured for five distinct pairs of fluorescent dyes\cite{Schuler2014}. 
In equilibrium conditions, namely when $[\mbox{ATP}]/[\mbox{ADP}] \ll 1$, the calculated FRET efficiency is $\simeq$ 0.8 for all considered pairs of fluorescent dyes (blue circles) and it matches the experimental results for the compact unbound rhodanese (~0.8). A dramatic difference is observed in excess of ATP (red circles), where the expansion of the substrate leads to a significant decrease of the calculated efficiency, in excellent agreement with the experimental values measured in similar conditions (black circles)\cite{Schuler2014}. Remarkably, the results correctly captured the non-monotonic behaviour of FRET efficiency as a function of the sequence separation between the dyes, which was not reproduced in previous calculations\cite{Schuler2014}. This agreement corroborates the prediction of the DnaK binding sites on the rhodanese sequence and the overall reliability of our model.

\subsection{Energetic balance and thermodynamic efficiency}

Molecular chaperones consume energy via ATP hydrolysis in order to expand rhodanese. It is hence important to determine how effective they are as molecular machines, as well as to assess how favourable the physiological conditions are to perform their biological task.

\begin{figure}
	\begin{center}
		\includegraphics[width=\linewidth]{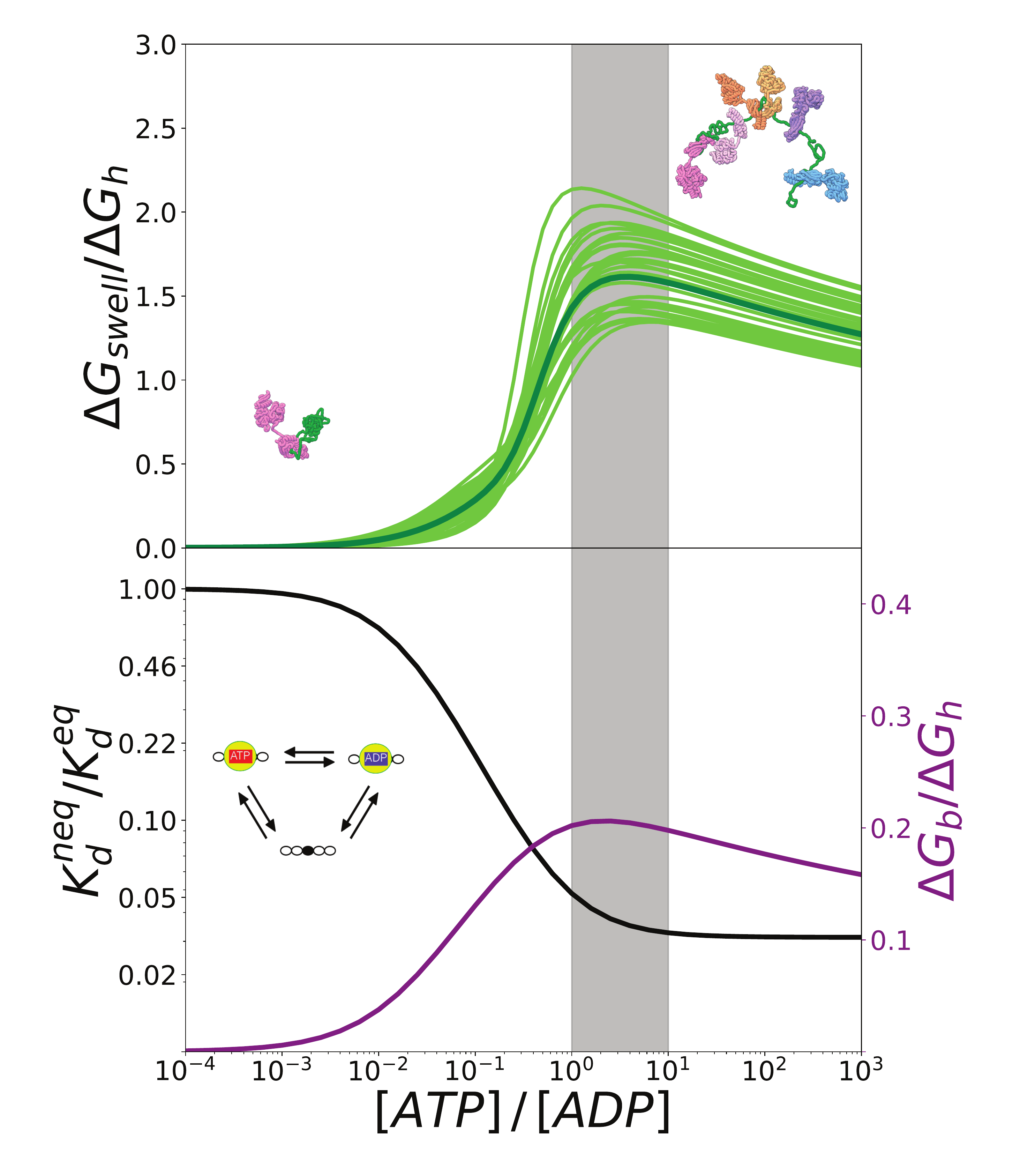}
		\caption{\label{fig:fig7} (top)  Ratio between the conformational free energy  and the free energy of ATP hydrolysis, as a function of [ATP]/[ADP]. Light green curves represent single realizations, whereas the dark green curve is the average. (bottom) Effective dissociation constant in the case of a single binding site normalized with respect to the corresponding value in equilibrium, as a function of [ATP]/[ADP] (black). Ratio between the binding free energy and the free energy of ATP, as a function of [ATP]/[ADP] (purple). The gray region indicates the interval of the physiological conditions.} 
	\end{center}
\end{figure}

To this aim, we calculated the global increase in the overall conformational free energy of the substrate  with respect to equilibrium conditions, $\Delta G_{Swell}$(Fig.\ref{fig:fig7}, top). This quantity measures the excess probability 
of each complex with respect to equilibrium conditions weighted by its corresponding conformational free-energy $\Delta G_i$.
\begin{equation}
\Delta G_{Swell}=\sum_{\mbox{i}}\left[p_i\left(\frac{[ATP]}{[ADP]}\right)-p^{eq}_i\right]\Delta G_i,
\end{equation} 
where $p_i\left(\frac{[ATP]}{[ADP]}\right)$ is the probability of complex $i$ for a given value of $[ATP]/[ADP]$ and $p^{eq}_i$ is the same quantity computed at equilibrium conditions. 
In order to investigate the conversion of chemical energy into mechanical work it is instructive to focus on the ratio between $\Delta G_{Swell}$ and the free energy of hydrolysis of ATP $\Delta G_{h}$, which reports on the effectiveness of the transduction process. We plot in Fig.\ref{fig:fig7} (top) this quantity as a function of the $[ATP]/[ADP]$ ratio considering the estimated inaccuracies of the model as previously done for the gyration radius. Not surprisingly, all these curves exhibit a maximum because the probabilities of the different states, and thus also $\Delta G_{swell}$, attain plateaus for $[ATP] \gg [ADP]$, whereas $\Delta G_{h}$ increases monotonically with the nucleotide ratio (see Methods). The maximal transduction regime intriguingly corresponds to values of $[ATP]/[ADP]$ that are typical of cellular conditions (grey area). 

We highlight that in our model Hsp70 functioning encompasses two distinct yet intertwined processes: the ATP-dependent binding of the chaperones to the substrate, and its consequent expansion. 
Energy transduction occurs then through two steps, and the amount of energy available for the mechanical expansion is limited by that provided by chaperone binding.
To corroborate this picture and to gain a more complete insight on the action of Hsp70s, we  analyzed the energetic balance of chaperone binding to a single site. To this aim, we focused on a simplified reaction cycle, which essentially corresponds to a single triangle within the overall scheme in Fig.\ref{fig:scheme} and does not take into account the conformational free-energy of the substrate.
We report in Fig.\ref{fig:fig7} (lower panel, black) the non-equilibrium dissociation constant, $K^{neq}_d$, normalized with respect to its equilibrium value $K^{eq}_d$, as a function of [ATP]/[ADP]. When the ratio between the concentrations of ATP and ADP approaches the physiological regime, the dissociation constant drops significantly until it settles at a value that is two order of magnitude lower than its equilibrium counterpart (this result is the core of ultra-affinity). We can convert the dissociation constant into a binding free energy excess with respect to equilibrium
\begin{equation}
\Delta G_b= -k_B T \ln\left[\frac{K^{neq}_d}{K^{eq}_d}\right]
\end{equation}
that we can compare to the free-energy of ATP hydrolysis, $\Delta G_{h}$, as previously done in the case of $\Delta G_{Swell}$. Interestingly, also in this case the energy ratio is maximal in cellular conditions,
suggesting that the optimality of the overall expansion process (Fig.\ref{fig:fig7}, top panel) does not depend on specific features of the substrate but it is a direct consequence of the intrinsic kinetic parameters of Hsp70 binding. 

\section{Discussion}

Integrating molecular simulations, polymer theory, single-molecule experimental data and non-equilibrium rate models, we have developed a comprehensive framework that provides a quantitative picture of Hsp70-induced expansion of substrate proteins and offers a broad insight into the cellular functioning of this versatile chaperone machine.

We relied on molecular simulations for characterizing the structural and thermodynamic features of the complexes formed by the bacterial chaperone DnaK and its unfolded substrate rhodanese. Notably, we investigated a large variety of possible chaperone-substrate complexes for determining their conformational free-energy as a function of stoichiometry and chaperone binding patterns. This computational strategy based on an enhanced-sampling protocol confirmed that excluded volume interactions upon chaperone binding can greatly perturb the conformational ensemble of the unfolded substrate leading to its expansion. Remarkably, simulation results were found to be in excellent agreement with the predictions of Sanchez theory for globule to coil transition, thus providing another example of how polymer theory can be successfully used to decipher the behaviour of disordered proteins \cite{Schuler_dis1,Schuler_dis2,haran}. 
We then combined conformational free-energies with available biochemical data to develop an analytical rate model of the chaperone/substrate reaction cycle, which included both chaperone binding/unbinding and nucleotide hydrolysis/exchange processes. 

This model fully takes into account non-equilibrium effects due to ATP hydrolysis and represents a natural extension of the ultra-affinity framework originally developed for peptide substrates with a single Hsp70 binding site\cite{Bard_ultra}. We could thus investigate the population of each complex and the average structural properties of the system as a function of the ATP/ADP nucleotide ratio, which measures how far the system is from thermodynamic equilibrium. The reliability of the model was corroborated by quantitative comparison with recent sm-FRET data, indicating that our non-equilibrium framework accurately captures the salient features of the ATP-dependent expansion. We then used this unprecedented access to the thermodynamics details of this complex molecular process to compare the free-energy cost associated with substrate swelling with the chemical energy released by ATP-hydrolysis. Remarkably, this analysis revealed that energy transduction is maximally efficient for ATP/ADP values in cellular conditions. This result hints at the possibility that Hsp70 chaperones have been tuned by evolution to optimize the conversion of chemical energy into mechanical work for substrate expansion. Further analysis indicated that this optimality is likely inherited from the intrinsic properties of Hsp70 chaperones, which can convert up to 20\% of the ATP chemical energy into non-equilibrium, excess binding energy at physiological conditions (Fig.\ref{fig:fig7}, lower panel). 
  
Hsp70s are highly versatile machines that play a fundamental role in a variety of diverse cellular functions beyond the unfolding of non-native substrates, such as protein translocation, protein translation, and disassembly of protein complexes. Nevertheless, all these processes share basic analogies from the mechanistic point of view. Indeed, in all these cases Hsp70 binding to flexible substrates in constrained environments requires the energy of ATP hydrolysis (ultra-affinity) and results in the generation of effective forces due to excluded volume effects (entropic pulling), which ultimately drive protein translocation into mitochondria \cite{Delos_entr_pull2,assenza2015quantifying}, clathrin cage disassembly \cite{Sousa} and/or prevention of ribosome stalling \cite{Liu}. 
Here by detailing how energy flows from ATP hydrolysis to mechanical work due to entropic pulling, we have elucidated a general force-generating mechanism of Hsp70 chaperones. This mechanism does not rely on any power-stroke conformational change but it rather depends on the efficient conversion of ATP chemical energy into ultra-affinity. The non-equilibrium nature of this process allows further spatial and temporal regulation by cofactors, thus paving the way for performing complicated molecular functions such as the ATP-dependent stabilization of native proteins \cite{Goloubinoff2018}.

\section{Materials and methods}
\subsection{Details of the MD simulations}
In our MD simulations both rhodanese and Hsp70 were coarse grained at the single-residue level. To this aim, the molecules were represented as collections of beads centered on the $C_\alpha$ atom of each amino acid. 
Rhodanese was modeled according to the force field from Ref. \cite{smith2014calibrated} (see SI for details). We modeled Hsp70 starting from the known crystal structure of ADP-bound Hsp70 (PDB:2KHO \cite{bertelsen2009solution}). The influence of Hsp70 in the ATP state on substrate conformation was assumed to give similar results. Following \cite{assenza2015quantifying}, we considered both the NBD and the SBD to be rigid bodies interacting only via excluded-volume interactions, while we modeled the flexible linker in the same way as rhodanese. The residues of the binding site moved rigidly with the corresponding SBD, thus ensuring that each chaperone was irreversibly bound to the substrate. 
All the simulations were performed with a version of LAMMPS \cite{lammps} patched with PLUMED \cite{plumed}. The temperature $T=293\,\mbox{K}$ was controlled through a Langevin thermostat with damping parameter $16 \,\mbox{ns}^{-1}$. The time step was set  equal to 1~fs, and each residue had a mass equal to 1 Da to speed up equilibration. In equilibrium simulations, the system was equilibrated for $10^7$ time steps starting from a rod-like conformation, and subsequently sampled for other $10^7$ time steps. At least 10 independent simulations were performed for each configuration. Statistical errors on the computed quantities were estimated as standard errors of the mean. In the pulling simulations, starting from an equilibrium conformation the substrate was pulled by an external force acting on its radius of gyration $R_g$ at a constant pulling speed $v=10^{-5}$ \AA$/$fs, until a rod-like conformation was reached, which we defined to be at $R_g=260$ \AA. For this set of simulations, 100 independent realizations for each configuration were performed. The indetermination of the computed free energies were estimated according to the bootstrap method. For both equilibrium and pulling simulations, the statistical errors are smaller than the size of symbols reported in the figures.\\
  
  \subsection{Rate model} 
  For the rate model we considered all possible binding configurations for chaperones in the ATP or ADP state to the six identified binding sites, resulting in a total of $3^6 = 729$ states. The equation for the average concentration of each binding configuration $c_i$ has the form
  \begin{equation}
    \frac{dc_i}{dt}=\sum_j k_{ji}c_j - \sum_j k_{ij} c_i\,,
\label{rate equations}
  \end{equation}
where $k_{ij}$ is the transition rate from configuration $i$ to $j$. The transition rates between different states depend on the intrinsic rates of the underlying molecular processes, namely binding and unbinding of a chaperone from a given binding side, the hydrolysis and synthesis of the nucleotides, and the nucleotide exchange, and have been taken according to biochemistry experiments  \cite{Schuler2014,Tomita,MayerBukau2000} (see SI for the rate values and for details about the thermodynamic constraints between the rates). 

We considered the steady state solution, obtained by imposing 
\begin{equation}
    \frac{dc_i}{dt}=0\,,
\end{equation}
for every $i$.
The concentration of chaperones was fixed at $[\mbox{Hsp70}]=10\,\mu M$, and the concentration of nucleotides was equal to $1\, \mbox{mM}$ as in the experiments in \cite{Schuler2014}. 
The free energy provided by the hydrolysis of ATP is given by the following formula:
  \begin{equation}
      \Delta G_h=k_bT\left[\ln\left(\frac{[ATP]}{[ADP]}\right)-\ln\left(\frac{[ATP]_{eq}}{[ADP]_{eq}}\right)\right]\,,
  \end{equation}
  where $[ATP]_{eq}$ and $[ADP]_{eq}$ are the concentrations of ATP and ADP in equilibrium.
  The mathematical and physical details of the model, as well as the values of the rates, are described in the Supplementary Information.

\subsection{Molecular graphics}
Molecular graphics in Figs.\ref{fig:histoRg_DG} and \ref{fig:fig7} have been generated with UCSF Chimera, developed by the Resource for Biocomputing, Visualization, and Informatics at the University of California, San Francisco, with support from NIH P41-GM103311 \cite{Chimera}.

\end{document}


\newcommand{\rates}[2]{\displaystyle \mathrel{\longrightleftharpoons^{#1\mathstrut}_{#2}}}

\begin{center}
 \textbf{\large Supplementary Information for ``Efficient conversion of chemical energy into mechanical work by the Hsp70 chaperones''}
\vspace{0.2cm}

S. Assenza$^{1,2}$, A. S. Sassi$^3$, R. Kellner$^4$, B. Schuler$^{4,5}$, P. De Los Rios$^{3,6}$ and A. Barducci$^7$

\vspace{0.2cm}
{\small $^1$ Laboratory of Food and Soft Materials, ETH Z\"urich, CH-8092 Z\"urich, Switzerland\\
$^2$ Departamento de F\'isica Te\'orica de la Materia Condensada, Universidad Aut\'onoma de Madrid, E-28049 Madrid, Spain\\
$^3$ Institute of Physics, School of Basic Sciences, Ecole Polytechnique F\' ed\' erale de Lausanne (EPFL), CH-1015 Lausanne, Switzerland\\
$^4$ IBM T. J. Watson Research Center, Yorktown Heights, New York, United States of America\\
$^5$ Department of Biochemistry, University of Zurich, CH-8057 Zurich, Switzerland\\
$^6$ Department of Physics, University of Zurich, CH-8057 Zurich, Switzerland\\
$^7$ Institute of Bioengineering, School of Life Sciences, Ecole Polytechnique F\'ed\'erale de Lausanne (EPFL), CH-1015 Lausanne, Switzerland\\
$^8$Inserm, U1054, Montpellier, France}
\end{center}

\vspace{1cm}
\textbf{\large Details of MD Simulations}
\vspace{0.2cm}

{\bf Force field.} Following \cite{smith2014calibrated},  two- and three-body bonded interactions along the substrate were included via harmonic potentials, namely $V_{\mbox{\tiny bond}}=k_l\sum_b\left(r_{b}-l\right)^2/2$ and $V_{\mbox{\tiny bend}}=\frac{1}{2}k_\theta\sum_\alpha\left(\theta_\alpha-\theta_0\right)^2$, respectively. In the previous formulas, $r_b$ denotes bond lengths; $\theta_\alpha$ the bend angles; $l=3.9$ \AA; $(k_l/k_BT)^{-1/2}=0.046$ \AA; $\theta_0=2.12$ rad; $(k_\theta/k_BT)^{-\frac{1}{2}}=0.26$; and $k_BT$ is the thermal energy. Four-body bonded interactions were implemented as Fourier terms, $V_{\mbox{\tiny dihed}}=k_BT\sum_d\sum_{s=1}^4
 \left[A_s\cos\left(s\phi_d\right)+B_s\sin\left(s\phi_d\right)\right]$, where $\phi_d$ is the torsion angle and $A_1=0.705$, $A_2=-0.313$, $A_3=-0.079$, $A_4=0.041$, 
$B_1=-0.175$, $B_2=-0.093$, $B_3=0.030$, $B_4=0.030$.
The steric repulsion was implemented through a Weeks-Chandler-Andersen potential, $V_{WCA}=\sum_{ij}V_r$, where 
\begin{equation}
V_r=\begin{cases} 
      4k_BT\left[\left(\frac{\sigma}{r_{ij}}\right)^{12}-\left(\frac{\sigma}{r_{ij}}\right)^6\right]
      +k_BT & \;\mbox{if }r_{ij}\leq 2^{\frac{1}{6}}\sigma \\
      0 & \;\mbox{\small otherwise} \\ 
   \end{cases}\,.
\end{equation}
In the previous formula, $r_{ij}$ is the distance between beads $i$ and $j$, while $\sigma=4.8$ \AA. The hydrophobic part of the potential is specific to the interacting residues and is modeled as the attractive part of the Lennard-Jones potential,
$V_{\mbox{\tiny hydro}}=\epsilon_h\sum_{ij}V_h$, where
\begin{equation}
 V_h=\begin{cases} 
      4\epsilon_{ij}\left[\left(\frac{\sigma}{r_{ij}}\right)^{12}-\left(\frac{\sigma}{r_{ij}}\right)^6\right]
      & \;\mbox{if }r_{ij}\geq 2^{\frac{1}{6}}\sigma \\
      -\epsilon_{ij} & \;\mbox{\small otherwise} \\ 
   \end{cases}\,.
\end{equation}
In the previous formula, $\epsilon_h=0.7722\,k_BT$ sets the overall 
strength of the hydrophobic interactions, while $\epsilon_{ij}$ depends 
on the residues $i$ and $j$ involved in the interaction, and is defined as the 
geometric mean of their hydrophobicities, 
$\epsilon_{ij}\equiv\sqrt{\epsilon_{i}\epsilon_{j}}$. The values of the hydrophobicities considered are based on a shifted and normalized Monera hydrophobicity scale\cite{smith2014calibrated}. Without further tuning, this force field gives a radius of gyration of unbound rhodanese equal to $R_g=(23.3\pm 0.1)$ \AA$\,$, which is in good agreement with the experimental value $R_g=(20.1\pm 0.8)$ \AA$\,$ \cite{kellner2014single}.

{\bf Binding sites.} DnaK binding sites were selected by applying the algorithms by van Durme et al. \cite{vanDurme} and R\"udiger et al. \cite{Rudiger}. Only the sequences were chosen for which at least partial consensus between the two approaches was present. In this way, six binding sites were identified centered at the following amino acids along rhodanese: 10, 118, 131, 162, 188, 260. 

\begin{figure}[h!]
\begin{center}
\includegraphics[width=0.5\textwidth]{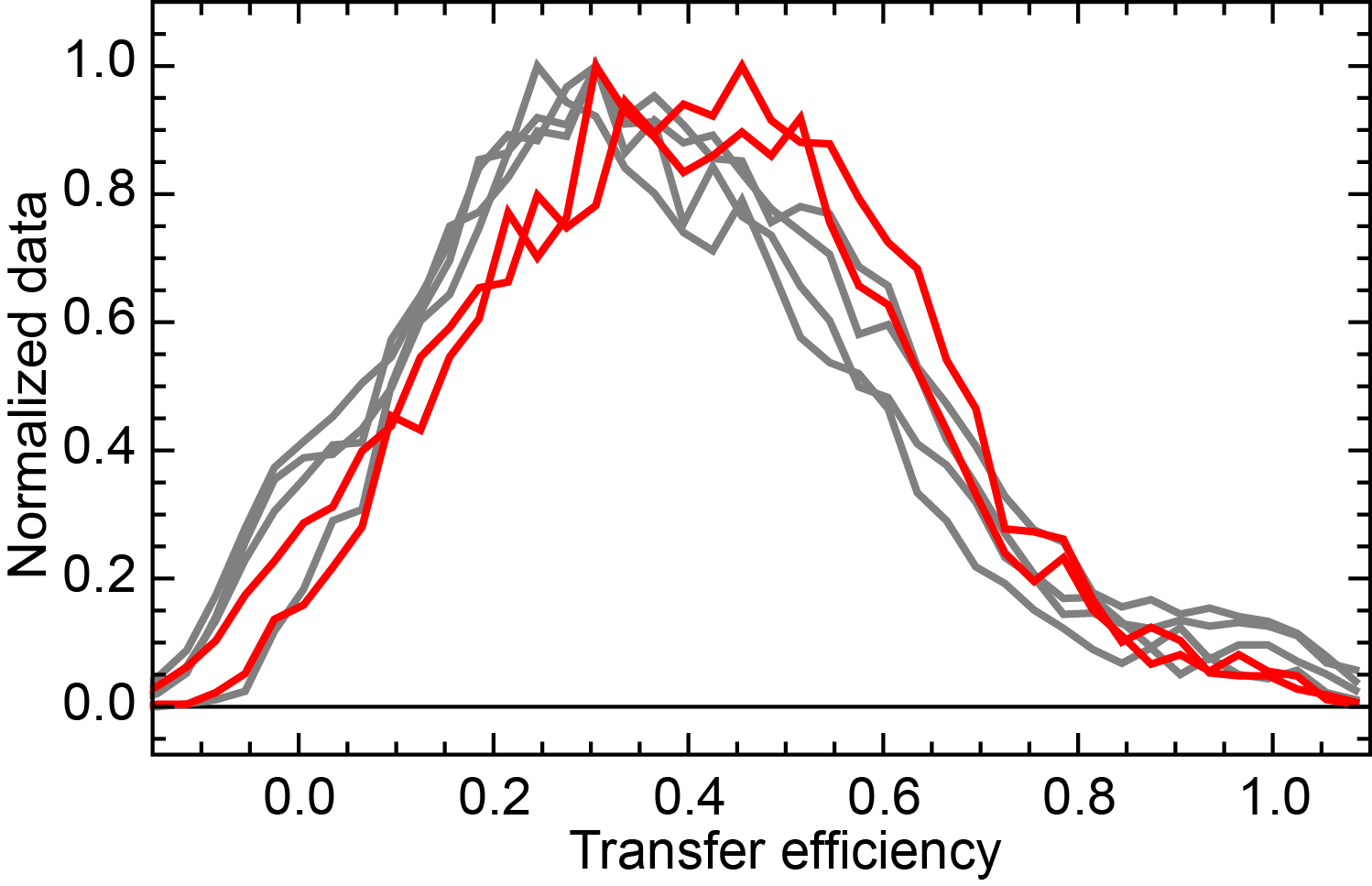}
\caption{\label{fig:ruthelectro} Normalized FRET efficiency histograms of the variant $\Delta39$ (K135C/K174C). Experiments were performed in the presence of 10 $\mu$M DnaK, 500 nM DnaJ, 1 mM ATP and either 5 mM (grey curves) or 100 mM (red curves) of KCl.}
\end{center}
\end{figure}

{\bf Experimental assessment of electrostatic interactions.} The choice of neglecting electrostatic interactions in the CG simulation of DnaK/rhodanese complexes was motivated by the results of two control experiments, which were performed exactly at the same conditions except for the salt concentration, namely 5 mM and 100 mM KCl. A double-cysteine variant of rhodanese (K135C/K174C), which was produced by site-directed mutagenesis and prepared as described before for wild-type rhodanese \cite{miller1992recombinant}, was labeled with Alexa Fluor 488 C5 maleimide and Alexa Fluor 594 C5 maleimide (Invitrogen, Molecular Probes) \cite{hillger2007detection}. A twofold molar excess of the dyes was added to the protein and incubated for 1h at room temperature. Unreacted dye was removed by gel filtration followed by anion exchange chromatography to reduce the amount of incorrectly labelled protein using a MonoQ 5/50 GL column (GE Healthcare) installed on an \"AKTA purifier FPLC system equilibrated in 50 mM Tris·HCl, pH 7.0, and eluted with a gradient from 0 to 500 mM sodium chloride over 60 mL (12 column volumes). The chaperone proteins DnaK and DnaJ (stock solution concentration 100 $\mu$M in 50 mM Tris HCl, pH 7.7, 100 mM NaCl) were gifts from H.-J. Sch\"onfeld (Hoffmann-La Roche Ltd., Basel). Labelled rhodanese was denatured in 4 M guanidinium chloride in buffer (50 mM Tris HCl, 10 mM MgCl$_2$, 200 mM $\beta$-mercaptoethanol, and 0.001\% Tween 20) with either 5 or 100 mM KCl added. The denatured rhodanese was diluted 100x into buffer containing 10 $\mu$M DnaK, 0.5 $\mu$M DnaJ, 1 mM ATP and 5 or 100 mM KCl to form chaperone rhodanese complexes at a final concentration of ~50 pM rhodanese. Single-molecule F\"orster Resonance Energy Transfer (FRET) measurements were started immediately after dilution and data recorded at 22$^\circ$C for 30 min to construct the FRET efficiency histograms. 
Data were recorded with a MicroTime 200 confocal microscope (PicoQuant) and on a custom-built confocal microscope. All measurements were obtained with pulsed interleaved excitation \cite{muller2005pulsed}. The instrument set up and data
reduction were the same as described before \cite{kellner2014single}.\\
The collected FRET histograms are plotted in Fig.\ref{fig:ruthelectro}. The average FRET efficiencies were $0.38 \pm 0.01$ with 5 mM KCl and $0.40 \pm 0.01$ with 100 mM KCl (errors indicate uncertainties estimated from the standard deviation of the larger data set collected at 5 mM salt). The very similar values obtained at different salt concentrations suggest that electrostatic interactions do not play a significant role in determining the conformational properties of rhodanese/chaperone complexes.

{\bf Control simulations on electrostatic interactions.} To further test the role of electrostatic for the results of the present work, a direct comparison was performed for the case of DnaK-free rhodanese by considering a set of simulations including also
electrostatics interactions following Smith et al.\cite{smith2014calibrated}. 
Similar values of the radius of gyration were found for simulations with and without electrostatics, namely $R_g=(22.5\pm 0.1)$ \AA\hspace{0.01cm} and 
$R_g=(23.3\pm 0.1)$ \AA\hspace{0.01cm} respectively.
These results are in line with the analysis performed in the work by Smith et al.\cite{smith2014calibrated} indicating that electrostatic interactions affects the  conformational ensembles only in the case of highly charged proteins.

\begin{figure}[h!]
\begin{center}
\includegraphics[width=0.7\textwidth]{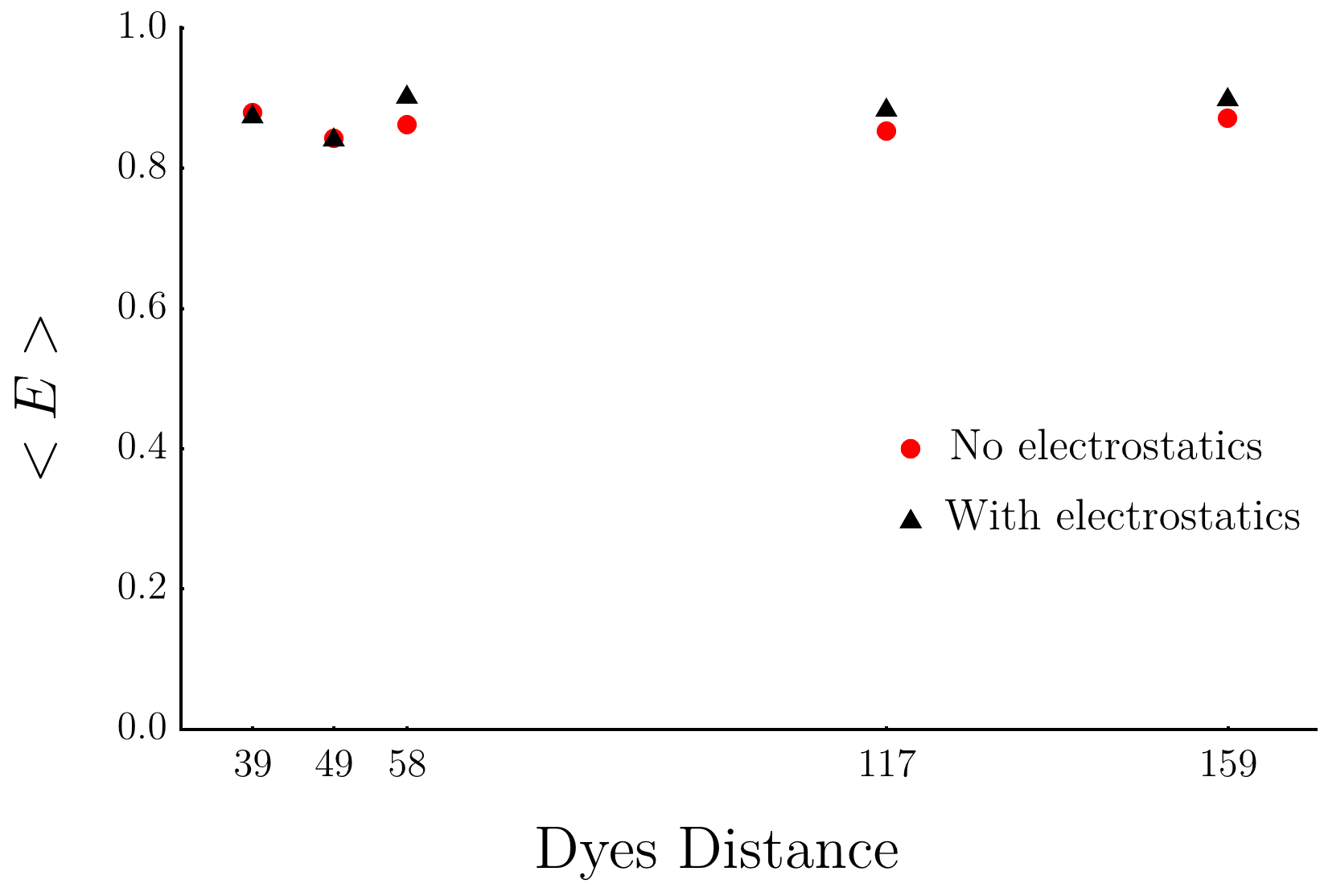}
\caption{\label{fig:efficiency} Average FRET efficiency $<E>$ obtained from simulations in the a bsence of chaperones for the various couples of dyes considered in Ref.\cite{kellner2014single}. Black triangles were obtained considering the whole force field (excluded volume + hydrophobic +
electrostatic non-bound interacions), while for red circles electrostatic interactions were neglected.}
\end{center}
\end{figure}

\begin{figure}[h!]
\begin{center}
\includegraphics[width=0.7\textwidth]{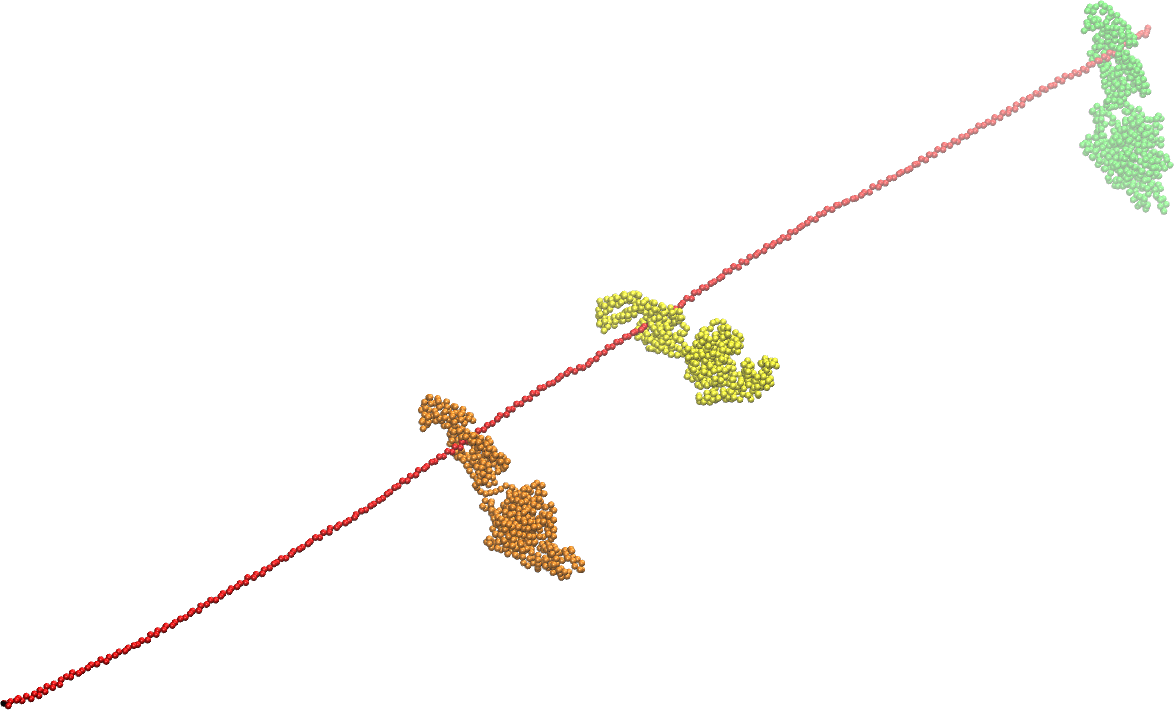}
\caption{\label{fig:stecco} Example of ``fully-stretched'' conformation with three bound chaperones.}
\end{center}
\end{figure}

{\bf Pulling simulations.} The dependence of conformational free energy of rhodanese on the set of bound chaperones was assessed via steered Molecular Dynamics simulations.
For any given chaperone configuration, the system was started at a conformation compatible with equilibrium and pulled by adding a harmonic potential
acting on the radius of gyration $R_g$ of rhodanese. The center of the harmonic trap was increased at a constant pulling speed 
$v=10^{-5}$\AA$/$fs from the equilibrium value up to an almost fully stretched conformation, which was set at 
$R_g=R_g^{\mbox{\tiny fin}}\equiv$290 \AA\hspace{0.01cm} (Fig.\ref{fig:stecco}).\\
\begin{figure}[h!]
\begin{center}
\includegraphics[height=0.85\textheight]{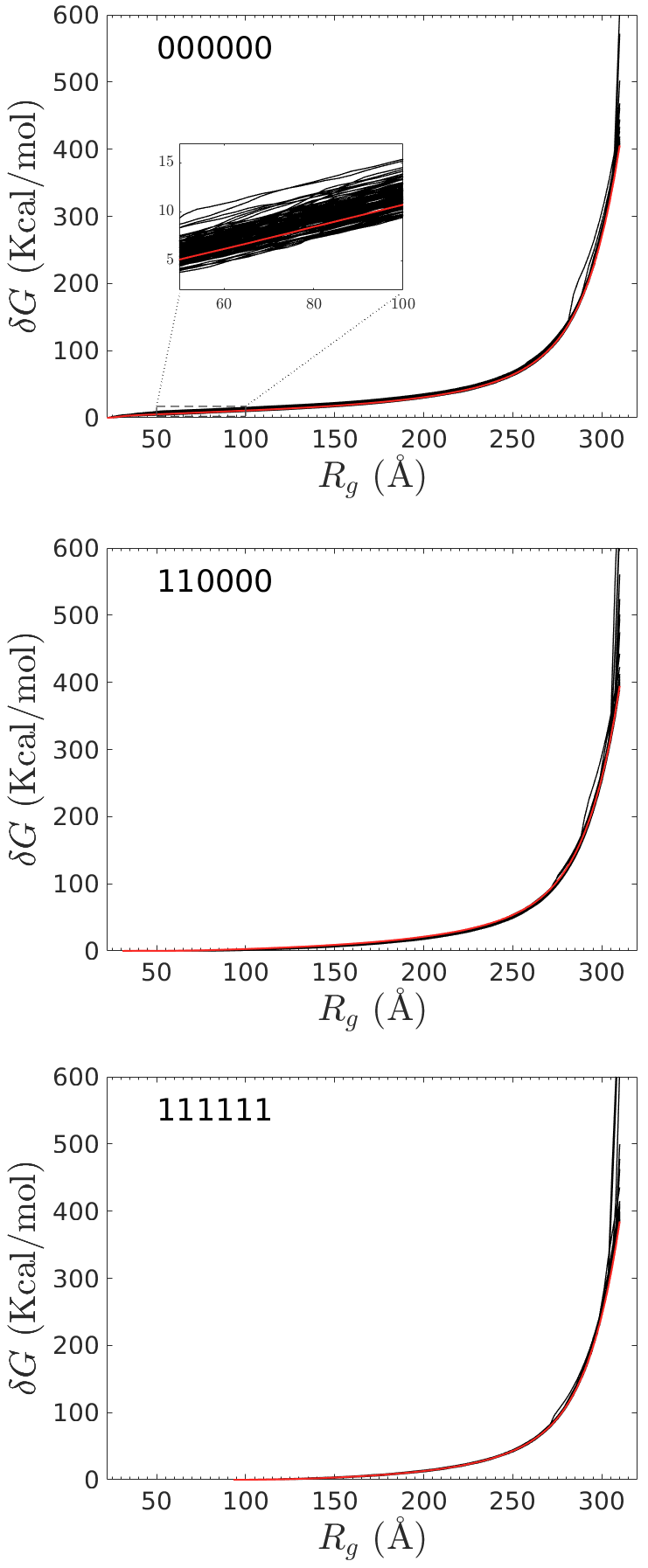}
\caption{\label{fig:pullingexamples} Results from pulling simulations for representative configurations with 0 (top), 2 (middle)
and 6 (bottom) bound chaperones. The particular configuration is identified by a string of six figures corresponding to the six 
binding sites, with ``1'' and ``0'' indicating bound and unbound chaperone, respectively. For instance, in the middle panel 
the first and second binding sites are occupied by chaperones. Black curves indicate the work $W$ performed by the bias 
potential in each realization, while the red curves give the averages obtained according to the Jarzynski equality.}
\end{center}
\end{figure}
Free-energy differences between the various configurations of bound chaperones and the case of simple rhodanese (i.e., no bound Hsp70s) were obtained
via the Jarzynski equality \cite{jarzynski1997nonequilibrium} in the following way. For each combination of bound chaperones, 100 independent
pulling realizations were performed, starting from uncorrelated initial snapshots extracted from the equilibrium distribution. For each realization, the work $W$
performed by the bias potential during the pulling process was measured (black curves in Fig. \ref{fig:pullingexamples}). 
The Jarzynski equality then reads \cite{jarzynski1997nonequilibrium}
\begin{equation}
 e^{-\frac{\delta G}{k_BT}}=\left<e^{-\frac{W}{k_BT}}\right>\,,
 \label{eqn:jarzynski}
\end{equation}
where $\delta G$ is the free-energy difference between the equilibrium starting point and the state corresponding to the chosen value of $R_g$, while $\left<\dots\right>$
denotes statistical average. Equation (\ref{eqn:jarzynski}) enables the computation of $\delta G(R_g)$ 
from the knowledge of the pulling work, leading to the red curves reported in Fig.\ref{fig:pullingexamples}.
Errors were estimated according to the bootstrap method.\\
Due to the large intermolecular distances, the effect of chaperones on 
the conformational properties of fully-stretched rhodanese is negligible (Fig.\ref{fig:stecco}).
Hence, this state can be taken as a reference for free-energy computations. The Jarzynski averages were consequently shifted so that 
$\delta G(R_g^{\mbox{\tiny fin}})=0$ for all the chaperone combinations. As an example, in Fig.\ref{fig:compare} the averages considered in 
Fig.\ref{fig:pullingexamples} are shifted according to this prescription. After such adjustment, the vertical distance between 
the starting point of each pulling curve and the case
with no bound chaperones (arrows in Fig.\ref{fig:compare}) gives the estimation 
for the free-energy difference $\Delta G$ considered in the main text. In order to enhance the robustness 
of the results, the final values reported in the main text were obtained as a further average over the values of $R_g^{\mbox{\tiny fin}}$ within the 
range $260$\AA\hspace{0.01cm}$\leq R_g^{\mbox{\tiny fin}}\leq 290$\AA\hspace{0.01cm}.

\begin{figure}[h!]
\begin{center}
\includegraphics[width=0.85\textwidth]{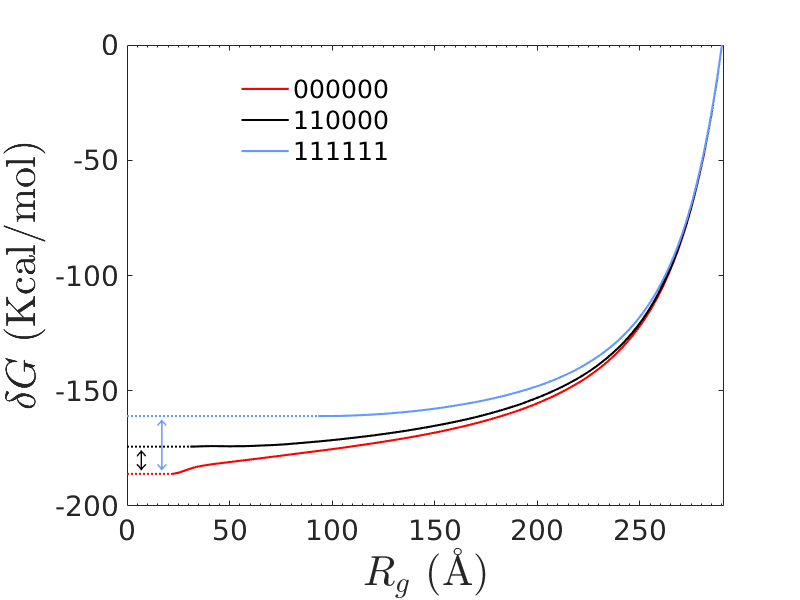}
\caption{\label{fig:compare} Shifted free energies $\delta G$ for the same chaperone configurations as in Fig.\ref{fig:pullingexamples}}.
\end{center}
\end{figure}

\newpage
\vspace{1cm}
\textbf{\large Sanchez theory}
\vspace{0.2cm}

In his theory on polymer coil-to-globule transition, Sanchez \cite{Sanchez} considers a Freely-Jointed Chain (FJC) made of $n$ monomers characterized by  attractive interactions of average magnitude $\epsilon$. Let $\alpha\equiv R_g/R_{g,0}$ be the expansion parameter, where $R_g$ is the radius of gyration of the polymer, while $R_{g,0}$ is the value of $R_g$ in the unperturbed case ($\epsilon=0$ and no excluded volume present). Assuming a Flory-Fisk distribution for the unperturbed case, the probability distribution of the expansion parameter $P(\alpha)$ was shown to be \cite{Sanchez}
\begin{equation}
P(\alpha)=\frac{1}{Z}\alpha^6 e^{-\frac{7}{2}\alpha^2+nq(\epsilon,\alpha)}\,,   
\label{eqn:sanchez}
\end{equation}
where $Z$ is the partition function and
\begin{equation*}
q(\epsilon,\alpha)=\frac{1}{2}\epsilon\frac{\phi_0}{\alpha^3}-
\left(\frac{\alpha^3}{\phi_0}-1\right)\ln \left(1-\frac{\phi_0}{\alpha^3}\right)\,,
\end{equation*}
with $\phi_0=\sqrt{19/(27n)}$. The free energy can then be straightforwardly computed 
as $\Delta G(\alpha)=-k_BT\ln P(\alpha)$, i.e., making the dependence on $R_g$ explicit,
\begin{equation}
\Delta G(R_g)=-k_BT\ln P(R_g/R_{g,0})\,.
\label{eqn:DeltaGSanchez}
\end{equation}

In the present case, rhodanese is constituted by $293$ amino acids, i.e. there are $N_b=292$ bonds. The bond length is $b_l=3.8$ \AA, estimated as the typical distance between $C_\alpha$ atoms belonging to consecutive
residues \cite{hofmann}. In order to apply the Sanchez theory, one needs to consider the FJC equivalent to rhodanese. To this aim, the Kuhn length $b_K$ can be estimated as $b_K=2l_p=8$ \AA, where $l_p=4$ \AA$\,$ is 
the persistence length of a protein \cite{hofmann}. The number of monomers $n$ in the equivalent FJC can then
be computed by imposing the total contour length of rhodanese: $n = N_b b_l/b_K$. From there, both 
$\phi_0$ and the unperturbed radius of gyration $R_{g,0}=\sqrt{n b_K^2/6}$ can be calculated, leaving
$\epsilon$ as the only unknown quantity in Eq. (\ref{eqn:sanchez}). The latter can be fixed by imposing
that $\Delta G$ has a minimum at the experimental value $R_g^*=20.1$\AA$\,$ obtained
in the absence of chaperones \cite{kellner2014single}, yielding $\epsilon\simeq 1.07$ kcal/mol.

Plugging the values of the parameters derived above into Eq.(\ref{eqn:DeltaGSanchez}) gives
the black line reported in Fig.2 in the main text, in excellent agreement with the simulation results.
We stress that neither the force field nor the MD results were employed in the derivation of the 
parameters used in the theoretical formula.

\vspace{1cm}
\textbf{\large Computation of Efficiency}
\vspace{0.2cm}

The FRET efficiency $E$ for a given couple of dyes was computed starting from the distance $r$ separating 
the corresponding amino acids as \cite{kellner2014single}
\begin{equation}
 E=\frac{1}{1+\left(\frac{r}{r_0}\right)^6}\,,
\end{equation}
where $r_0=54 $\AA\hspace{0.01cm}. For each realization, the time average of $E$ was computed. The final values employed to compute the results reported in the inset of Fig.5 in the main text were obtained as the mean between independent realizations. The corresponding indeterminacy 
was estimated considering the standard error of the mean and computing the error propagation to the final values, and is smaller than the size of symbols.

 \vspace{1cm}
\textbf{\large Rate model}
\vspace{0.2cm}\\
For the kinetic model we consider a system in which each binding site can either be occupied by a chaperone in the ATP or ADP state, or it can be free, so that in total there are $3^6=729$ possible configurations. The concentration $c_i$ of each state  evolves in time according to a system of rate equations 
\begin{equation}
    \frac{dc_i}{dt}=\sum_j k_{ji}c_j - \sum_j k_{ij} c_i\,
\label{rate equations}
\end{equation}
where $k_{ij}$ is the transition rate from state $i$ to state $j$. 
The first term in the right hand side (\textit{r.h.s.}) of (\ref{rate equations}) represents the total flux of molecules from the other states toward state $i$, while the second term in the r.h.s. of (\ref{rate equations}) accounts for the flux of molecules from state $i$ to any other state.
We focused on the steady-state, when the concentrations of the various states do not change over time, which is defined by 
\begin{equation}
    \frac{dc_i}{dt}=0
\end{equation}\\
Here we provide a pedagogical example of state-encoding, and of the associated transitions and equations.
For example, in the configuration $(0,T,0,D,0,0)$, the first, third, fifth and sixth Hsp70 binding sites are unoccupied, the second binding site is associated to chaperone in the ATP-bound state while the fourth binding site is associated with a chaperone in the ADP-bound state. With this notation, the rates corresponding to every reaction are easily determined. 
The main reactions that need to be considered are 
\begin{itemize}
    \item binding/unbinding 
    \begin{align*}
        (...0...) &\ce{<=>C[k^{adp}_{on}\,e^{-\beta\Delta\Delta G}][{k^{adp}_{off}}]} (...D...)\\
        (...0...) &\ce{<=>C[k^{atp}_{on}\,e^{-\beta\Delta\Delta G}][{k^{atp}_{off}}]} (...T...)
    \end{align*}
    \item hydrolysis/synthesis
    \begin{align*}
        (...D...) &\ce{<=>C[k^s_s][{k^s_h}]} (...T...)\\
    \end{align*}
    \item nucleotide exchange
    \begin{align*}
        (...D...) &\ce{<=>C[k^{eff}_{ex,DT}][{k^{eff}_{ex,TD}}]} (...T...)\,.
    \end{align*}
    
\end{itemize}
We further provide, as an example, the equation for a precise configuration, say $(0,T,0,D,0,0)$. The two binding sites that are occupied can undergo chaperone unbinding, ATP hydrolysis/synthesis or nucleotide exchange. 
The remaining unoccupied binding sites can bind either an ATP- or an ADP-bound chaperone. We thus have
\begin{align}
    \frac{d}{dt}(0,T,0,D,0,0)=&-(0,T,0,D,0,0)*(k^{eff}_{ex,DT} + k^s_s+ k^{eff}_{ex,TD}+k^s_h+k^{atp}_{off}+k^{adp}_{off})+\\
    &+(0,0,0,D,0,0)[\mbox{Hsp70}\cdot \mbox{ATP}]k^{atp}_{on}e^{-\beta\Delta\Delta G}+\nonumber\\
    &+(0,T,0,0,0,0)[\mbox{Hsp70}\cdot \mbox{ADP}]k^{adp}_{on}e^{-\beta\Delta\Delta G}+\nonumber\\
    &+(T,T,0,D,0,0)k^{atp}_{off}+\nonumber\\
    &+(D,T,0,D,0,0)k^{adp}_{off}+\nonumber\\
    &+(0,T,T,D,0,0)k^{atp}_{off}+\nonumber\\
    &+(0,T,D,D,0,0)k^{adp}_{off}+\nonumber\\
    &+(0,T,0,D,T,0)k^{atp}_{off}+\nonumber\\
    &+(0,T,0,D,D,0)k^{adp}_{off}+\nonumber\\
    &+(0,T,0,D,0,T)k^{atp}_{off}+\nonumber\\
    &+(0,T,0,D,0,D)k^{adp}_{off}\,.\nonumber
\end{align}
\\

Here below we further detail the rates of our model.

It is possible to move from an ATP-state to an ADP-state either via hydrolysis/synthesis or via nucleotide exchange. In the case of exchange, effective constants are used, which take into account the unbinding of one nucleotide species and the binding of the different one. The effective exchange rates are thus a function of the ratio [ATP]/[ADP] (see also \cite{bard_ultra}):
\begin{align}
    k^{eff}_{ex,DT}&=\frac{k_{-D}k_{+T}\frac{[ATP]}{[ADP]}}{k_{+D}+k_{+T}\frac{[ATP]}{[ADP]}}\\
        k^{eff}_{ex,TD}&=\frac{k_{-T}k_{+D}}{k_{+D}+k_{+T}\frac{[ATP]}{[ADP]}}\,,
\end{align}
where $k_{+D}$, $k_{+T}$, $k_{-D}$ and $k_{-T}$ are the binding and unbinding rates for ADP and ATP respectively. 

The rates of binding between the chaperone and single peptides have been previously determined experimentally \cite{MayerBukau2000}, and they were corrected in order to take into account the conformational change of the full polypeptide substrate upon binding, as we illustrated in the main text.  

Substrate binding enhances the chaperone ATPase activity. Furthermore, the stimulation of the hydrolysis of ATP always takes place in cooperation with the co-chaperone JDP. In our model we did not consider it explicitly but its contribution was implicitly included through the choice of the rate constants. 

If we call $k_h$ the hydrolysis rate in the absence of the substrate and $k^s_h$ the rate in the presence of the substrate, we have $k_h \ll k^s_h$.  
We assumed that the ratio between the rate of hydrolysis $k_h$ and the rate of synthesis $k_s$ is not altered by the substrate:
\begin{equation}
    \frac{k_h}{k_s}=\frac{k^s_h}{k^s_s}\,.
\end{equation}

The substrate binding/unbinding rates, the rates of nucleotide exchange and the hydrolysis and synthesis rates are collectively constrained by thermodynamic relations.
Indeed, when the ratio between the concentrations of ATP and ADP is equal its equilibrium value (when the spontaneous hydrolysis and synthesis reaction are at steady state and compensate each other), detailed balance must be satisfied \cite{Ge}. As a consequence, for every closed cycle in the reaction network the product of the rates in one direction must be equal to the product of the rates in the opposite direction.
Therefore, if $k^{atp}_{on}$, $k^{adp}_{on}$, $k^{atp}_{off}$ and $k^{adp}_{off}$ are the rate of substrate binding and unbinding from a chaperone in the ATP and ADP states, we must have 
\begin{equation}\label{product}
   \frac{ k^{atp}_{on}k^s_hk^{adp}_{off}k_s}{k^{adp}_{on}k^s_s k^{atp}_{off}k_h}=\frac{ k^{atp}_{on}k^{adp}_{off}}{k^{adp}_{on}k^{atp}_{off}}=1\,.
\end{equation}.

Remarkably, taking the rates as provided in \cite{kellner2014single,MayerBukau2000,MayerHuTomita}, this relation is not satisfied, and we had thus to modify them.
We just calculated the product in the formula above and then corrected the rates in the following way:
\begin{align}
       \frac{ k^{atp}_{on}k^{adp}_{off}}{k^{adp}_{on}k^{atp}_{off}}&=r\\
       k^{atp}_{on}\,,\,k^{adp}_{off}&\rightarrow k^{atp}_{on}/r^{1/4}\,,\,k^{adp}_{off}/r^{1/4}\\
       k^{adp}_{on}\,,\,k^{atp}_{off}&\rightarrow k^{adp}_{on}*r^{1/4}\,,\,k^{atp}_{off}*r^{1/4}\,.
\end{align}

The concentration of free chaperones in the ATP and in the ADP states was obtained, at the leading order, by solving a three-state system whose reactions have the form
\begin{equation}
  \mbox{Hsp70}+\mbox{ADP} \ce{<=>}  \mbox{Hsp70}\cdot\mbox{ADP} \ce{<=>} \mbox{Hsp70}\cdot\mbox{ATP} \ce{<=>} \mbox{Hsp70}+\mbox{ATP} \,.
\end{equation}
Since we worked in the assumption of excess of chaperones in the system, once these concentrations were obtained, they remained fixed once for all, without being considered as a variable of the biochemical network. \\

We report in the following table the rates used in the model. 
\begin{center}
 \begin{tabular}{|c c | c c|} 
 \hline
  \multicolumn{4}{|c|}{{\bf Parameters of the model} \cite{kellner2014single,MayerBukau2000,MayerHuTomita}} \\
 \hline
 $k^{atp}_{off}$ & $2.31\,s^{-1}$ & $k^{adp}_{off}$ & $2*10^{-3}\,s^{-1}$ \\ 
 \hline
  $k_{-T}$ & $1.33*10^{-4}\,s^{-1}$ & $k_{-D}$ & $0.022\,s^{-1}$ \\
\hline
 $k^{atp}_{on}$  & $1.28*10^6\,M^{-1}s^{-1}$ & $k^{adp}_{on}$  & $10^3\,M^{-1}s^{-1}$ \\
 \hline
 $k_{+T}$ & $1.3*10^5\,M^{-1}s^{-1}$ & $k_{+D}$ & $2.67*10^5\,M^{-1}s^{-1}$ \\
 \hline
 $k_h$ & $6*10^{-4}\,s^{-1}$ &  $k^s_h$ &  $1.8\,s^{-1}$ \\ [1ex] 
 \hline
\end{tabular}
\end{center}

To test the robustness of the model for the radius of gyration, the average FRET efficiency and the free energy $\Delta G_{swell}$, 100 realizations were implemented, taking each time the values $\Delta G_i$ from a Gaussian distribution with $\sigma=0.3$ kcal/mol and mean equal to the value obtained with the MD simulations.\\
\vspace{1cm}

\renewcommand\bibname{References}